\documentclass[twocolappendix,twocolumn,tighten,times]{aastex631}

\usepackage{relsize}
\usepackage{subfigure}

\shorttitle{The Radio Galaxy/Quasar Dipole}
\shortauthors{Secrest et al.}

\begin{document}

\title{A Challenge to the Standard Cosmological Model}

\correspondingauthor{Nathan J.\ Secrest}
\email{nathan.j.secrest.civ@us.navy.mil}

\author[0000-0002-4902-8077]{Nathan J.\ Secrest}
\affiliation{U.S.\ Naval Observatory, 3450 Massachusetts Ave NW, Washington, DC 20392-5420, USA}

\author[0000-0002-6274-1424]{Sebastian von Hausegger}
\affiliation{Rudolf Peierls Centre for Theoretical Physics, University of Oxford, Parks Road, Oxford, OX1 3PU, United Kingdom}

\author[0000-0001-5023-5631]{Mohamed Rameez}
\affiliation{Dept. of High Energy Physics, Tata Institute of Fundamental Research, Homi Bhabha Road, Mumbai 400005, India}

\author[0000-0002-5944-3995]{Roya Mohayaee}
\affiliation{Rudolf Peierls Centre for Theoretical Physics, University of Oxford, Parks Road, Oxford, OX1 3PU, United Kingdom}
\affiliation{Sorbonne Universit\'e, CNRS, Institut d'Astrophysique de Paris, 98bis Bld Arago, Paris 75014, France}

\author[0000-0002-3542-858X]{Subir Sarkar}
\affiliation{Rudolf Peierls Centre for Theoretical Physics, University of Oxford, Parks Road, Oxford, OX1 3PU, United Kingdom}

\begin{abstract}
We present the first joint analysis of catalogs of radio galaxies and quasars to determine if their sky distribution is consistent with the standard $\Lambda$CDM model of cosmology. This model is based on the cosmological principle, which asserts that the universe is statistically isotropic and homogeneous on large scales, so the observed dipole anisotropy in the cosmic microwave background (CMB) must be attributed to our local peculiar motion. We test the null hypothesis that there is a dipole anisotropy in the sky distribution of radio galaxies and quasars consistent with the motion inferred from the CMB, as is expected for cosmologically distant sources. Our two samples, constructed respectively from the NRAO VLA Sky Survey and the Wide-field Infrared Survey Explorer, are systematically independent and have no shared objects. 
Using a completely general statistic that accounts for correlation between the found dipole amplitude and its directional offset from the CMB dipole, the null hypothesis is independently rejected by the radio galaxy and quasar samples with $p$-value of $8.9\times10^{-3}$ and $1.2\times10^{-5}$, respectively, corresponding to $2.6\sigma$ and $4.4\sigma$ significance. 
The joint significance, using sample size-weighted $Z$-scores, is $5.1\sigma$. We show that the radio galaxy and quasar dipoles are consistent with each other and find no evidence for any frequency dependence of the amplitude. 
The consistency of the two dipoles improves if we boost to the CMB frame assuming its dipole to be fully kinematic, suggesting that cosmologically distant radio galaxies and quasars may have an intrinsic anisotropy in this frame.
\end{abstract}

\section{Introduction} \label{sec:intro}
The $\Lambda$CDM cosmological model is based on the isotropic and homogeneous Friedmann-Lema\^{i}tre-Robertson-Walker (FLRW) metric, justified by the ``cosmological principle'' that the universe on large scales must appear to be the same to all observers, independent of their location \citep{1937RSPSA.158..324M}. According to the standard picture of the growth of structure through gravitational instability, the distribution of matter on cosmological scales (larger than $\sim 100\,h^{-1}$~Mpc) reflects the linear evolution of primordial adiabatic density perturbations of amplitude $\sim 10^{-5}$ seen imprinted on the cosmic microwave background (CMB).\footnote{$H_0 \equiv 100\,h$~km\,s$^{-1}$Mpc$^{-1}$ is the present expansion rate, with $h \simeq 0.7$.} The large-scale distribution of matter must therefore share the same ``cosmic rest frame'' as the CMB (also called the ``CMB frame'') in which the Friedmann-Lema\^{i}tre equations hold. As observed from the solar system barycentric frame, however, the CMB exhibits a prominent dipole anisotropy with amplitude $\mathcal{D} \sim 10^{-3}$, which is attributed to our peculiar velocity with respect to this frame. This motion is routinely corrected for in, for example, analysis of the Hubble diagram of Type~Ia supernovae, and in estimation of the $\Lambda$CDM model parameters from CMB data.

The cosmological principle has served well as a simplifying assumption that made quantitative cosmology possible. The inference that the universe is dominated by a cosmological constant $\Lambda$ of ${\cal O}(H_0^2)$, however, rests crucially on this assumption \citep[see, e.g.,][]{2008GReGr..40..269S}; hence it is essential to rigorously test it. The most direct way of doing so is to check if the distribution of matter on cosmological scales is indeed isotropic in the CMB frame. A model-independent method for doing this was proposed by \citet{1984MNRAS.206..377E}: consider an observer moving at velocity $v\ll c$ with respect to an isotropic distribution of distant sources. Within the observer's instrumental passband, the sources have a  power-law spectral energy distribution of the form $S \propto \nu^{-\alpha}$, and the apparent flux density $S$ of the sources within this passband has a cumulative power-law distribution $N(>S) \propto S^{-x}$. If the observer surveys the sky down to a specific flux density above which the completeness of the survey is unbiased with respect to direction, then relativistic aberration and Doppler boosting of source emission in the observer's frame will induce a dipole anisotropy in the sky distribution of sources with amplitude \citep{1984MNRAS.206..377E}: 

\begin{equation}
\mathcal{D} = \left[ 2 + x (1+\alpha) \right] \beta ,
\label{eq:D}
\end{equation}
\noindent 
where $\beta \equiv v/c$. This is called the kinematic dipole, and the null hypothesis is that the direction and amplitude of the dipole of distant matter matches the direction of the CMB dipole and the velocity inferred from its amplitude.\footnote{Throughout this work, we use $v=369.82 \pm 0.11$~km\,s$^{-1}$ towards $(l,b)=(264\fdg021,48\fdg253)$ \citep{2020A&A...641A...1P}.} 

The minimum number of sources required to perform this test is of ${\cal O}(10^5)$ \citep{1984MNRAS.206..377E}, which precluded statistically significant  constraints until the advent of the 1.4~GHz NRAO~VLA~Sky~Survey \citep[NVSS;][]{1998AJ....115.1693C}. This enabled the creation of large samples of radio galaxies down to a completeness limit of $\sim10$~mJy, yielding several estimates of the radio galaxy dipole \citep{2002Natur.416..150B, 2011ApJ...742L..23S, 2012MNRAS.427.1994G, 2013A&A...555A.117R, 2015MNRAS.447.2658T, 2017MNRAS.471.1045C}. All of these studies found the radio dipole to be over a factor of 2 
larger than the kinematic expectation, albeit with modest significance ($\sim2-3\sigma$). The anomalously large NVSS dipole has consequently been controversial, with some authors arguing that it is due to unidentified systematics in the data \citep[e.g.,][]{2012MNRAS.427.1994G} or possibly a large bias factor for radio galaxies at low redshift \citep[e.g.,][]{2016JCAP...03..062T}.
Complicating matters further, the 150~MHz TIFR GMRT Sky Survey \citep[TGSS;][]{2017A&A...598A..78I} appears to exhibit an even larger dipole \citep[e.g.,][]{2018JCAP...04..031B}, motivating a recent claim that the anomalous radio galaxy dipole is frequency dependent \citep{2021A&A...653A...9S}.

Confirmation of an anomalously large dipole of distant matter requires using data that is systematically independent --- not sharing the same instruments, survey design, or calibration method. This was  accomplished by \citet{2021ApJ...908L..51S}, who used quasars selected with mid-IR photometry from the Wide-field Infrared Survey Explorer \citep[WISE;][]{2010AJ....140.1868W}. The unique power of WISE data to reliably select large, nearly all-sky samples of quasars based on photometry alone was demonstrated by \citet{2015ApJS..221...12S}, and the release of the CatWISE2020 catalog \citep{2021ApJS..253....8M}, which contains much deeper photometry than the previous AllWISE release, enabled the construction of a cosmology-grade quasar catalog of 1.36~million objects. These have a mean redshift of $\langle z \rangle = 1.2$, with 99\% having $z>0.1$, thus precluding a significant contribution from the local clustering dipole. \citet{2021ApJ...908L..51S} found that the quasar dipole amplitude and direction, while similar to the previous results from NVSS, reject the null hypothesis with much higher statistical significance ($p=5\times10^{-7}$, or $4.9\sigma$). 

To date, however, the dipoles of radio galaxies and quasars have not been jointly analyzed. There are several important motivations to do this. First, the methodology used to determine the significance of disagreement with the expected kinematic dipole (e.g., treatment of survey systematics and estimation of errors) varies considerably in the literature, so a meta-analysis of published results can be misleading. Second, there is some overlap in the populations of radio galaxies and quasars that introduces correlation between results, motivating an analysis that explicitly removes shared sources. Third, a joint analysis may reveal a consistent, common amplitude and direction for the radio galaxy and quasar dipoles, which could be an important clue for cosmology.

In this Letter, we perform the first joint analysis of the sky distributions of distant radio galaxies and quasars, which independently provide the strongest constraints on the kinematic dipole of distant matter. In Section~\ref{sec:data}, we carefully account for survey systematics, such as declination-dependent sensitivity differences, as well as astrophysical systematics, such as Galactic synchrotron emission and reddening, all of which can introduce dependencies of source density on position. We assess the overlap between the radio galaxy and quasar populations, and account for shared sources to produce completely independent samples. In order to remain conservative and account for correlation between dipole positions and amplitudes, we also introduce a two dimensional, completely generalized $p$-value to assess the null hypothesis. Our results are given in Section~\ref{sec:results}, wherein we also explore if there is a dipole shared by the radio galaxy and quasar populations. In the Appendix, we critically address related results published in the literature since the publication of \citet{2021ApJ...908L..51S}, such as a possible frequency dependence of the anomalously large dipole \citep{2021A&A...653A...9S}, a considerably larger dipole found in an older catalog of active galactic nuclei (AGNs) from WISE \citep{2021Univ....7..107S}, and a recent result \citep{2022ApJ...931L..14D} claiming consistency between the radio galaxy dipole and the kinematic expectation. Our conclusions are presented in Section~\ref{sec:conclusion}.

\section{Galaxy Samples} \label{sec:data}
In this work, we use radio galaxies from the NVSS and quasars selected using mid-infrared photometry from CatWISE2020. The former is composed of radio galaxies detected in 1.4\,GHz continuum imaging taken with the Very~Large~Array (VLA) in New Mexico, while the latter is composed of quasars detected in $3.4\,\micron$ ($W1$) and $4.6\,\micron$ ($W2$) imaging taken with WISE, in a polar low Earth orbit. Being ground based, position-dependent systematics present in the NVSS depend on declination. Specifically, the NVSS used the compact VLA~D configuration for $-10\arcdeg<\mathrm{decl.}<+78\arcdeg$ and the hybrid DnC configuration for $-40\arcdeg<\mathrm{decl.}<-10\arcdeg$ and $\mathrm{decl.}>+78\arcdeg$. The NVSS images are composed of individual pointings mosaicked together on a hexagonal grid, following lines of right ascension, and adjusting declination spacing to account for increasing overlap at high latitude. The WISE scanning pattern, on the other hand, is aligned with the ecliptic for Sun avoidance, scanning the sky continuously in great circles that converge at the ecliptic poles, using a scan mirror to compensate for the telescope's motion during integrations. The single exposure images are mosaicked onto a grid of 18,240 predefined tiles shared across the various WISE data releases, with CatWISE2020 being the latest.

While the NVSS and WISE catalogs are independent, the systematics present in each must nonetheless be carefully addressed. This is done primarily by developing sky masks to mitigate instrumental systematics such as source confusion, image artifacts, and survey footprint limitations, as well as astrophysical systematics such as diffuse Galactic synchrotron that may affect the purity and uniformity of an extragalactic source catalog. Additionally, each catalog has an effective sensitivity limit, generally set by a position-dependent survey depth, which must be controlled for. In the following sections, we discuss the masks and flux density cuts developed for the NVSS and WISE catalogs. We use the Hierarchical Equal Area isoLatitude Pixelization \citep[HEALPix;][]{2005ApJ...622..759G}\footnote{\url{https://healpix.sourceforge.io}} scheme to bin dipole-subtracted source density by declination, ecliptic latitude, Galactic synchrotron emission, and other systematics of interest to ensure that the source density of the masked catalog does not show any trends with these systematics. Bin sizes are chosen to be large enough to calculate the reduced $\chi^2$ and reduce the statistical dispersion, but small enough that any trends present are not under-sampled. The reduced $\chi^2$ is defined as:

\begin{equation} \label{eq: rchi2}
\chi^2 / \mathrm{df} = \frac{1}{N - k} \sum_i^N \frac{(\rho_i - f_i)^2}{\sigma_{\rho_i}^2 / n_i}
\end{equation}

\noindent where $\rho_i$ is the mean source density in bin $i$, $f_i$ is the value of the functional fit for that bin, $\sigma_{\rho_i}$ is the dispersion of $\rho$ within bin $i$, $n_i$ is the number of sky pixels in the bin, and $k$ is the number of parameters corresponding to $f_i$. For example, the linear model fit with respect to ecliptic latitude used to de-trend the WISE sample (Section~\ref{sub:catwise}) has $k=2$. In checking the residuals of source density, $\rho$ is replaced by the residuals after subtraction of the dipole and monopole, so $f_i=0$ and $k=1$. We find that requiring 200 pixels per bin for $N_\mathrm{side}=64$ is a good compromise, although our results are not sensitive to changes in bin counts, and remain consistent if we use uniform bins and allow the number of pixels per bin to vary. We test for systematic trends in declination, ecliptic latitude, Galactic dust reddening, diffuse Galactic synchrotron emission, Galactic latitude, and supergalactic latitude. We use the \citet{2014A&A...571A..11P} map for dust, and the de-striped and source-subtracted version of the \citet{1982A&AS...47....1H} 408~MHz all-sky map made by \citet{2015MNRAS.451.4311R} for synchrotron emission.



\subsection{NVSS} \label{sub:nvss}
Using the full NVSS catalog, we identify highly localized ($\sim1\arcdeg$ scales) source concentrations that we use to produce a list of circular mask regions, setting the radii to fully encompass the concentration. As expected, these regions are generally within a few degrees of the Galactic plane, although some regions at high Galactic latitude were also identified, which are likely image artifacts near particularly bright radio sources such as M87. For less distinct artifact concentrations near the Galactic center, we use the diffuse synchrotron map and mask all pixels with a mean brightness temperature of 50\,K or higher. In total, 27\% of the sky was masked.

\begin{figure*}
    \includegraphics[width=\textwidth]{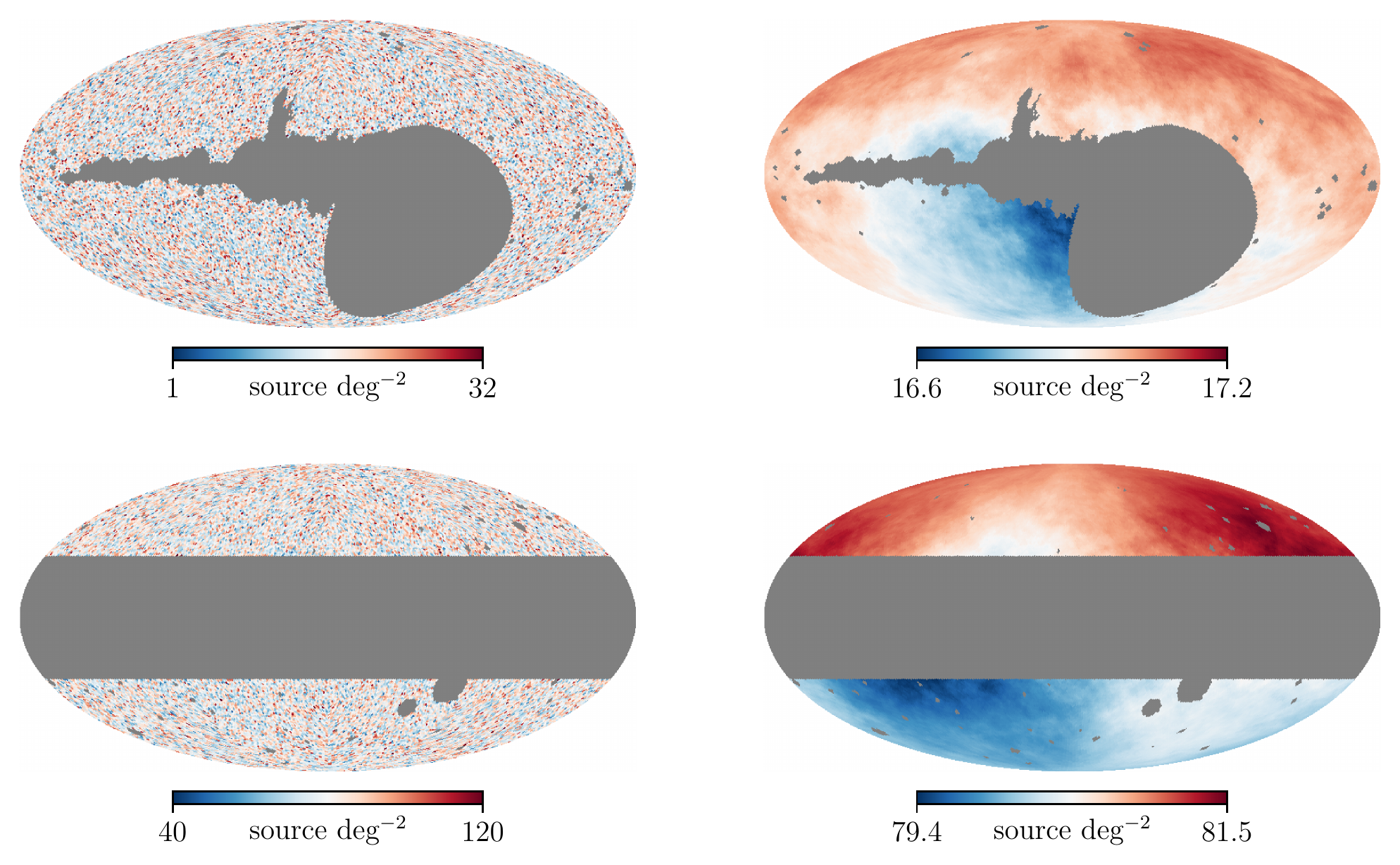}
    \caption{Top: Density map of the NVSS-based radio galaxy sample used in this work, in Galactic coordinates, Mollweide projection. The right plot is the smoothed map, using a 1~rad moving average, showing the underlying dipole signal. Bottom: Corresponding maps of the WISE-based quasar sample used in this work. The smoothed maps are only for visual purposes and were not used in our analysis.}
    \label{fig:moll}
\end{figure*}

With the masking complete, the next step is to determine the flux density cut. Using simulated point sources, \citet{1998AJ....115.1693C} determined the $\sim100\%$ source completeness limit of the NVSS to be 4~mJy. Using this flux density cut, however, there may still be some residual declination dependence of the catalog sensitivity, although at low statistical significance ($\chi^2/\mathrm{df}=1.4$). Cutting at 10~mJy removes this potential systematic ($\chi^2/\mathrm{df}=1.1)$. We see no evidence for source density dependence on any of the potential systematics we tested, with $\chi^2/\mathrm{df}$ ranging from 0.95 to 1.3 for $E(B-V)$, Galactic synchrotron, or any of the principal latitudes (declination and ecliptic/Galactic/supergalactic latitude). This flux cut leaves 508,144 sources in the masked map. We show the NVSS sample in Figure~\ref{fig:moll}, top.

\subsection{WISE Quasars} \label{sub:catwise}
In \citet{2021ApJ...908L..51S}, we developed a mid-IR quasar sample from the CatWISE2020 catalog, which is deeper and more uniform than the AllWISE catalog because of inclusion of data from the NEOWISE Reactivation mission \citep{2011ApJ...731...53M, 2014ApJ...792...30M}. This catalog was built using the $W1-W2\geq0.8$ cut of \citet{2012ApJ...753...30S} that reliably picks AGN-dominated objects, and a flux cut of $W1<16.4$~mag for uniform sensitivity across the sky. Objects below an absolute Galactic latitude of $30\arcdeg$ were excluded because of the drop in sky pixel density due to source confusion near the Galactic plane. A slight inverse linear trend was also observed between ecliptic latitude and sky density, which is potentially attributable to two effects. First, deeper coverage near the ecliptic poles increases sensitivity to faint sources that, while excluded by our flux cut, may cause deblending issues with brighter sources and lead to a loss of completeness. Second, shallower coverage near the ecliptic equator may lead to AGNs slightly bluer than $W1-W2\geq0.8$ scattering red-ward due to photometric error, increasing apparent source density if bluer AGNs are more common, as is implied in Figure~2 of \citet{2021ApJ...908L..51S}. A detailed characterization of these effects is beyond the scope of this work; for our purposes it suffices that the ecliptic latitude trend is easy to correct for.

We retain the $|b| \geq 30\arcdeg$~Galactic plane cut used by \citet{2021ApJ...908L..51S}, as well as the source mask, but with a minor revision: we found that some of the sky areas masked in \citet{2021ApJ...908L..51S} were either not optimally centered on the region of interest (e.g., diffraction spikes around bright stars), or were over-masked (i.e., with too large a radius). We manually reevaluated every region outside the Galactic plane cut, of which there are 48 in the updated mask. Including the Galactic plane cut, 51\% of the sky was masked. Repeating the tests done on the NVSS sample, we find minimal unexplained variance in source density as a function of the principal latitudes or Galactic foregrounds, with $\chi^2/\mathrm{df}$ ranging from 0.8 to 1.4. In performing these tests, we found that the flux density cut used in \citet{2021ApJ...908L..51S} can safely be relaxed slightly to $W1 < 16.5$~mag \citep[$S > 0.078$~mJy; see Section~2 of][for how $W1$ magnitudes are converted to flux densities]{2021ApJ...908L..51S}.

We wish to determine the significance with which the NVSS and WISE dipoles independently reject the null hypothesis. These catalogs must therefore not contain the same objects. To this end, we match the full NVSS catalog to the full CatWISE2020 catalog using a $40\arcsec$ match tolerance, chosen for completeness given the astrometric uncertainties of NVSS that imply offsets of about $\sim5\arcsec$ on average. We find that $99.7\%$ of the NVSS sources have a counterpart in the full CatWISE2020 catalog, with 99\% of matches within $\sim20\arcsec$. Nonetheless, only $1.4\%$ of the WISE quasars are in the NVSS sample, likely because radio-selected AGNs tend to have low accretion rates \citep[e.g.,][]{2007ApJ...658..815S} and be hosted by luminous elliptical galaxies, while mid-IR AGNs are bolometrically dominant, and preferentially reside in bluer, less clustered galaxies \citep[e.g.,][]{2009ApJ...696..891H}. Indeed, using a sample of AGNs selected with the same WISE color cut we employ here and a catalog of VLA sources from the COSMOS field, \citet{2012ApJ...753...30S} find that only $\sim2\%$ of WISE-selected AGNs are radio-loud, consistent with what we find here. By performing a joint analysis on radio-selected and infrared-selected AGNs, we are therefore testing two almost entirely different populations of objects, each with its own host galaxy type and environment. We removed the small fraction of the quasar sample that have counterparts in the NVSS sample, and further removed random WISE quasars from regions of the sky not shared by the NVSS sample in order to maintain uniformity. This results in a total of 1.6~million WISE quasars, shown in Figure~\ref{fig:moll}, bottom. 

\subsection{Testing the Null Hypothesis} \label{subsec: null}
In \citet{2021ApJ...908L..51S}, we simulated the kinematic dipole by applying relativistic aberration and Doppler boosting to individual sources, expressed as directional vectors, which were then converted into HEALPix maps. In this work, we use a method of simulating the kinematic dipole directly in sky pixel space, which is computationally much more efficient and allows a wide range of statistical tests. Our method identifies the equal areas of the sky pixels as the differential solid angle $d\Omega$. Then, each sky pixel $m_i$ has a Doppler factor $\delta_i$:

\begin{equation}
\delta_i = \gamma (1 + \beta \cos\theta_i)
\end{equation}

\noindent where $\gamma \equiv (1 - \beta^2)^{-1/2}$, and $\theta_i$ is the angular offset of sky pixel $m_i$ from the velocity vector corresponding to the CMB frame. The expected number count within each pixel is enhanced by $\delta^2$ times the monopole $\mathcal{M}$, which is further boosted by source brightening, which goes as $\delta^{x (1 + \alpha)}$. Putting these together, the expected value of a sky pixel $m_i$ modified by relativistic aberration and Doppler boosting is:

\begin{equation} 
m_i = \delta^{2 + x (1 + \alpha)}_i \mathcal{M}
\label{eq: expectation}
\end{equation}
\noindent

\noindent Simulated skies are created by using the non-masked sky pixels $m_i$ as the expectation values for random sampling from a Poisson distribution (shot noise). 

Because variance in pixel counts due to relativistic aberration and Doppler boosting occurs at the flux limit of the catalog, the value of $\alpha$ and $x$ used should be the values at this flux limit. As has been noted for the NVSS previously \citep[e.g.,][]{2017MNRAS.471.1045C}, a single value of the power-law index $x$ is not sufficient to describe the integral source counts, being flatter at the faint end and becoming steeper at higher flux densities. We fit the faint end near the flux cut, finding $x=0.77$. As the NVSS was observed at a single frequency, we do not have $\alpha$ separately for each object. However, the population mean in each pixel is the relevant quantity and this is expected to be very near the typical synchrotron value $\alpha\sim0.75$. We tested the effect of allowing $\alpha$ to have a dispersion of 0.4, estimated from a match to the lower-frequency SUMSS catalog \citep{2003MNRAS.342.1117M}, with each pixel varying as $0.4\,m_i^{-1/2}$. We find that the effect of not knowing $\alpha$ for each individual source is negligible. For the WISE catalog we do have $\alpha$ for each source, with a mean value of 1.06 at the 0.078~mJy flux density limit, and we find that $x = 1.89$. For this sample we include the small uncertainty of the ecliptic latitude correction in the null sky simulations by dividing the expectation map by a permutation of the correction for each simulation, with each permutation being drawn from the fit covariance matrix of the correction. The best-fit correction is used in the fit, maintaining fidelity to counting statistics.

To fit dipoles, we used a modified version of the Healpy \texttt{fit\_dipole} function, which uses the linear algebra routines in NumPy. Our version optimizes memory usage to enable large Monte Carlo simulations to be run efficiently. The expectation value maps generated using Equation~\ref{eq: expectation} have fit dipoles with amplitudes in agreement with Equation~\ref{eq:D} \citep{1984MNRAS.206..377E}, which predicts $\mathcal{D}=0.41\times10^{-2}$ for NVSS and $\mathcal{D}=0.73\times10^{-2}$ for WISE. We quote formal uncertainties on dipole fit parameters by permuting the masked maps with shot noise, propagating the uncertainties of any additional terms such as the ecliptic latitude trend seen in WISE.

In \citet{2021ApJ...908L..51S}, the definition of the $p$-value was the fraction of simulated skies with dipole amplitudes exceeding the kinematic expectation and with directions within the offset between the CMB dipole and the found quasar dipole. This was motivated by the fact that simulated dipoles at larger offsets are more likely to have a significant contribution from the ``noise'' dipole, which can increase their amplitudes. However, the amplitude and offset of simulated skies are correlated, with higher amplitudes generally exhibiting smaller offsets, so a found dipole with smaller amplitude but larger offset could be equally as inconsistent with the null hypothesis as one with larger amplitude and smaller offset.

In this work, we therefore adopt a completely general definition of the $p$-value. Null sky simulations fill a 2-dimensional space in dipole amplitude and offset, allowing for an estimate of the joint probability distribution. The found dipole exists along a contour of equal probability density, and the $p$-value is the fraction of null skies outside of this contour. There will be a larger fraction of null skies meeting this criterion, so the $p$-value will be larger (less significant). Our generalized approach is therefore the most conservative.

\section{Results} \label{sec:results}
\begin{figure*}
    \includegraphics[width=\textwidth]{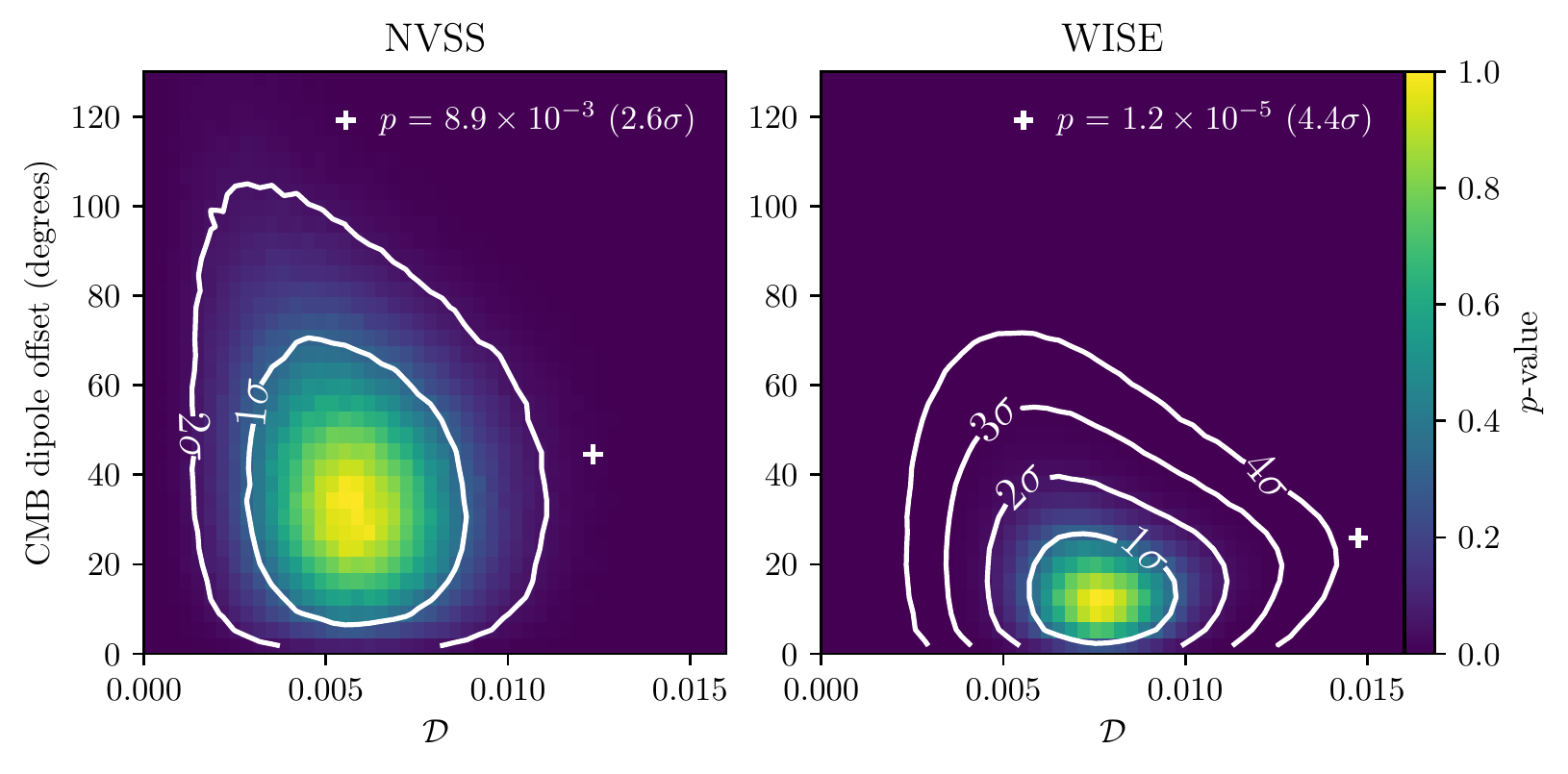}
    \caption{Distribution of CMB dipole offsets and kinematic dipole amplitudes of simulated null skies for the NVSS catalog (left) and WISE (right). Contours of equal $p$-value (scale on right y-axis), translated to equivalent $\sigma$ are given (where the peak of the distribution corresponds to $0\sigma$), with the found dipoles marked with the $+$ symbol and their $p$-value in the legends.}
    \label{fig:p_values}
\end{figure*}

We find an NVSS dipole amplitude of $\mathcal{D}=(1.23\pm0.25)\times10^{-2}$, exceeding the kinematic expectation by a factor of about 3, in the direction $(l,b)=(196\arcdeg\pm13\arcdeg,+46\arcdeg\pm10\arcdeg)$, $45\arcdeg$ away from the CMB dipole, with a 95\% upper confidence limit (CL) positional uncertainty of $30\arcdeg$. Testing the null hypothesis with $10^6$ simulated skies, we find it is rejected with a $p$-value of $8.9\times10^{-3}$, or $2.6\sigma$ (Figure~\ref{fig:p_values}, left). 
For WISE, we find $\mathcal{D}=(1.48\pm0.16)\times10^{-2}$, exceeding the kinematic expectation by a factor of about 2, in the direction $(l,b)=(238\arcdeg\pm7\arcdeg,+31\arcdeg\pm5\arcdeg)$, $26\arcdeg$ away from the CMB dipole, with a 95\% CL positional uncertainty of $15\arcdeg$. We performed $10^8$ null sky simulations, finding a $p$-value of $1.2\times10^{-5}$, which corresponds to $4.4\sigma$ (Figure~\ref{fig:p_values}, right). We note that the conversion from $p$-value to $\sigma$ is two-sided, so that the point of highest probability density corresponds to $0\sigma$.

Because each sample has its own particular mask, it is not straightforward to combine them to determine the joint significance with which the null hypothesis is rejected. Moreover, each sample has a different expected dipole amplitude under the null hypothesis, further complicating a single, combined test. However, the joint significance may be estimated using the weighted $Z$-score:

\begin{equation}
Z_\mathrm{joint} = \frac{\sum_i w_i Z_i}{\sqrt{\sum_i w_i^2}}
\end{equation}

\noindent 
where $w_i$ are the sample weights, in this case the square roots of the sample sizes (0.5 million for NVSS and 1.6 million for WISE), and $Z_i$ are the $Z$-scores of each sample independently, respectively $2.6$ and $4.4$. This gives $Z_\mathrm{joint} = 5.1$, or a joint significance of $5.1\sigma$ with which the kinematic expectation inferred from the CMB dipole in the standard cosmological model is rejected.

We note that, unlike in \citet{2021ApJ...908L..51S} we have not preserved the coupling of source fluxes and spectral indices, instead using the relevant values at the flux density limit of our catalog ($\alpha=1.06$, $x=1.89$). Repeating our methodology on the sample from \citet{2021ApJ...908L..51S} and defining the $p$-value in the same way, we get $p=3\times10^{-7}$, consistent with our previous result. This indicates that the effect of any correlation between $\alpha$ and $x$, as suggested by \cite{2022MNRAS.512.3895D}, is inconsequential for our results. There is likewise no evidence for a significant difference in $\alpha$ and $x$ between the hemispheres pointing towards the CMB dipole and  away from it, as would be expected in this scenario. The spectral index at the flux density limit is $\alpha = 1.06$ for both hemispheres in WISE, with uncertainties below the given precision. The values of $x$ in the towards/away hemispheres are 0.77/0.77 for NVSS, and 1.90/1.89 for WISE. The small difference in $x$ for WISE is consistent with fitting error, and  makes a negligible difference in the expected kinematic dipole amplitude.


As the dipoles in the large scale distribution of radio galaxies and of quasars independently reject the null hypothesis, we can ask if these two dipoles are consistent with each other and, if so, combine them to determine their common or shared dipole. We repeated the kinematic expectation test for a given input dipole amplitude and direction to determine the distribution in amplitude and offset. Using $10^6$ simulations, we find that the input dipole that is most consistent with the NVSS and WISE dipoles is their vector mean: $\mathcal{D}=(1.40\pm0.13)\times10^{-2}$, pointed at $(l,b)=(233\arcdeg\pm6\arcdeg,+34\arcdeg\pm5\arcdeg)$, $27\arcdeg$ offset from the CMB dipole, with a $14\arcdeg$ positional uncertainty at the 95\% CL. The corresponding $p$-value is $0.72$ for WISE and $0.09$ for NVSS, indicating that the NVSS and WISE dipoles are indeed consistent with each other, albeit with some tension in the NVSS sample. If we additionally assume that the CMB dipole is fully kinematic in origin, then the NVSS and WISE dipoles will each have a different kinematic contribution (with amplitudes $\mathcal{D}=0.41\times10^{-2}$ and $\mathcal{D}=0.73\times10^{-2}$, respectively), which can be removed from the samples using Equation~\ref{eq: expectation}. Doing this and repeating the above test yields a residual common dipole with amplitude $\mathcal{D}=(0.86\pm0.14)\times10^{-2}$, pointing towards $(l,b)=(217\arcdeg\pm10\arcdeg,+20\arcdeg\pm7\arcdeg)$, $48\arcdeg$ from the CMB dipole direction, with a 95\% CL position uncertainty of $22\arcdeg$. The corresponding $p$-values are $0.94$ for WISE and $0.30$ for NVSS, improving consistency and alleviating the tension with NVSS. This tantalizing result suggests that if the solar system barycenter is indeed traveling in the direction of the CMB dipole at $370$~km~s$^{-1}$, then the space distribution of cosmologically distant radio galaxies and quasars has an intrinsic dipole anisotropy in that frame. 

We reiterate that the two catalogs are completely independent of each other, not only systematically but also in terms of the objects they contain. The dipoles of radio galaxies and quasars are thus both larger than the kinematic expectation from the CMB dipole, but consistent with a common dipole which points $27\arcdeg$ away from the direction of the CMB dipole as observed, or $48\arcdeg$ away if the kinematic expectation is removed. Note that, according to  \citet{2022MNRAS.510.3098M}, the effect of gravitational lensing by the structures responsible for the local bulk flow is negligible for the dipole in cosmologically distant source counts.

Finally, since the NVSS and WISE samples were acquired at frequencies differing by nearly 5 orders of magnitude, their consistency disfavors any frequency dependence of the anomalous dipole as claimed by  \citet{2021A&A...653A...9S}. We discuss this claim in Appendix~\ref{app: tgss} and show that it can be attributed to known flux calibration issues in the 150~MHz TIFR GMRT Sky Survey catalog \citep[TGSS-ADR1][]{2017A&A...598A..78I}.

\section{Conclusions} \label{sec:conclusion}
We have explored the dipoles in the sky distributions of two large, independent, samples of radio galaxies and quasars, constructed from the NVSS and WISE catalogs. Our principal conclusions are as follows:

\begin{enumerate}
\item Using a common methodology and a completely generalized $p$-value, the large dipole anisotropies seen in radio galaxies and quasars independently reject, at $2.6\sigma$ and $4.4\sigma$ respectively, the null hypothesis that the dipoles arise due to Doppler boosting and relativistic aberration with velocity $370$~km~s$^{-1}$ in the direction of the CMB dipole. The found dipole amplitudes are about 3 and 2 times larger than the respective kinematic expectations, and point $45\arcdeg$ and $26\arcdeg$ away from the CMB dipole. The joint significance of this rejection of the cosmological principle is $5.1\sigma$.

\item These anomalously large dipoles are statistically consistent with a single, shared dipole of distant galaxies and quasars, with amplitude $\mathcal{D}=(1.40\pm0.13)\times10^{-2}$ in the direction $(l,b)=(233\arcdeg\pm6\arcdeg,+34\arcdeg\pm5\arcdeg)$. We find no evidence for a frequency dependence of the amplitude.

\item The agreement between the radio galaxy and quasar dipoles improves if the standard kinematic expectation is subtracted  out, yielding a dipole of amplitude $\mathcal{D}=(0.86\pm0.14)\times10^{-2}$ in the direction $(l,b)=(217\arcdeg\pm10\arcdeg,+20\arcdeg\pm7\arcdeg)$. This may be interpreted as an intrinsic over-density of galaxies and quasars on very large scales, in a direction $48\arcdeg$ away from the CMB dipole.

\end{enumerate}

These findings present a significant challenge to the cosmological principle and, by extension, the standard FLRW cosmological model. A better understanding of the anomalously large dipole of radio galaxies and quasars will require dedicated studies using data from ongoing surveys such as the Dark Energy Spectroscopic Instrument and the forthcoming Rubin Observatory Legacy Survey of Space and Time, as well as the Square Kilometre Array and the Euclid satellite. These data will enable the matter dipole to be traced as a function of redshift --- from $z \lesssim0.1$ where it can have a significant ``clustering dipole'' contribution from structure --- out to moderate $z$ where the kinematic dipole due to our local motion should prevail if the universe is indeed homogeneous and isotropic on large scales. Such tomographic studies may reveal if and how the observed anomalously large matter dipole is linked to our local bulk flow, which is also anomalous in extending deeper than is expected in the standard $\Lambda$CDM model of structure formation. Measurement of fluxes along with number counts will provide additional means to differentiate contributions to the matter dipole \citep{2015APh....61....1T,2021JCAP...11..009N}. 

\begin{acknowledgments}
We thank the anonymous referee for their helpful review of our paper. SvH acknowledges support from the Carlsberg Foundation.  The authors additionally thank Camille Bonvin, Enzo Branchini, Jacques Colin, Charles Dalang, Jim Peebles, Jean Souchay, and Jenny Wagner for helpful discussions. The National Radio Astronomy Observatory is a facility of the National Science Foundation operated under cooperative agreement by Associated Universities, Inc. This publication makes use of data products from the Wide-field Infrared Survey Explorer, which is a joint project of the University of California, Los Angeles, and the Jet Propulsion Laboratory/California Institute of Technology, funded by the National Aeronautics and Space Administration. Some of the results in this paper have been derived using the healpy and HEALPix package.
\end{acknowledgments}

\facilities{VLA, WISE}

\software{Astropy \citep{2013A&A...558A..33A, 2018AJ....156..123A},  dustmaps \citep{2018JOSS....3..695M},
          healpy \citep{2019JOSS....4.1298Z}, SciPy \citep{2020NatMe..17..261V}, \textsc{topcat} \citep{2005ASPC..347...29T}
          }

\appendix

\section{Related Results from the Literature}

Since the publication of \citet{2021ApJ...908L..51S}, a number of apparently discrepant results have appeared in the literature. For completeness, and to work towards concordance on this issue, we address the most salient findings here.

\subsection{TGSS} \label{app: tgss}
Several papers have examined the dipole in the TGSS, finding a significantly larger value than that of the NVSS. This is the main driver for the claim by \citet{2021A&A...653A...9S} that the radio dipole has a frequency dependence. Excluding the TGSS data point in their Figure~9, the WENSS, SUMSS and NVSS data points are all consistent with their weighted mean of $\langle \mathcal{D} \rangle=2.1\times10^{-2}$ with $\chi^2/\mathrm{df}=0.95$, an overall better fit than the functional dependence proposed by \citet{2021A&A...653A...9S} which has $\chi^2/\mathrm{df}=1.23$ after including the TGSS data point.

Indeed, the TGSS has two significant systematics that make this result suspect. First, it has significant, position-dependent flux calibration problems \citep[e.g.,][]{2017arXiv170306635H} that are visible in sky pixel maps across a range of flux density cuts. Second, it exhibits highly variable, position-dependent background RMS noise that correlates with source density \citep[see Figures~8 and B.2  in][]{2017A&A...598A..78I}. 

\begin{figure}
    \includegraphics[width=\columnwidth]{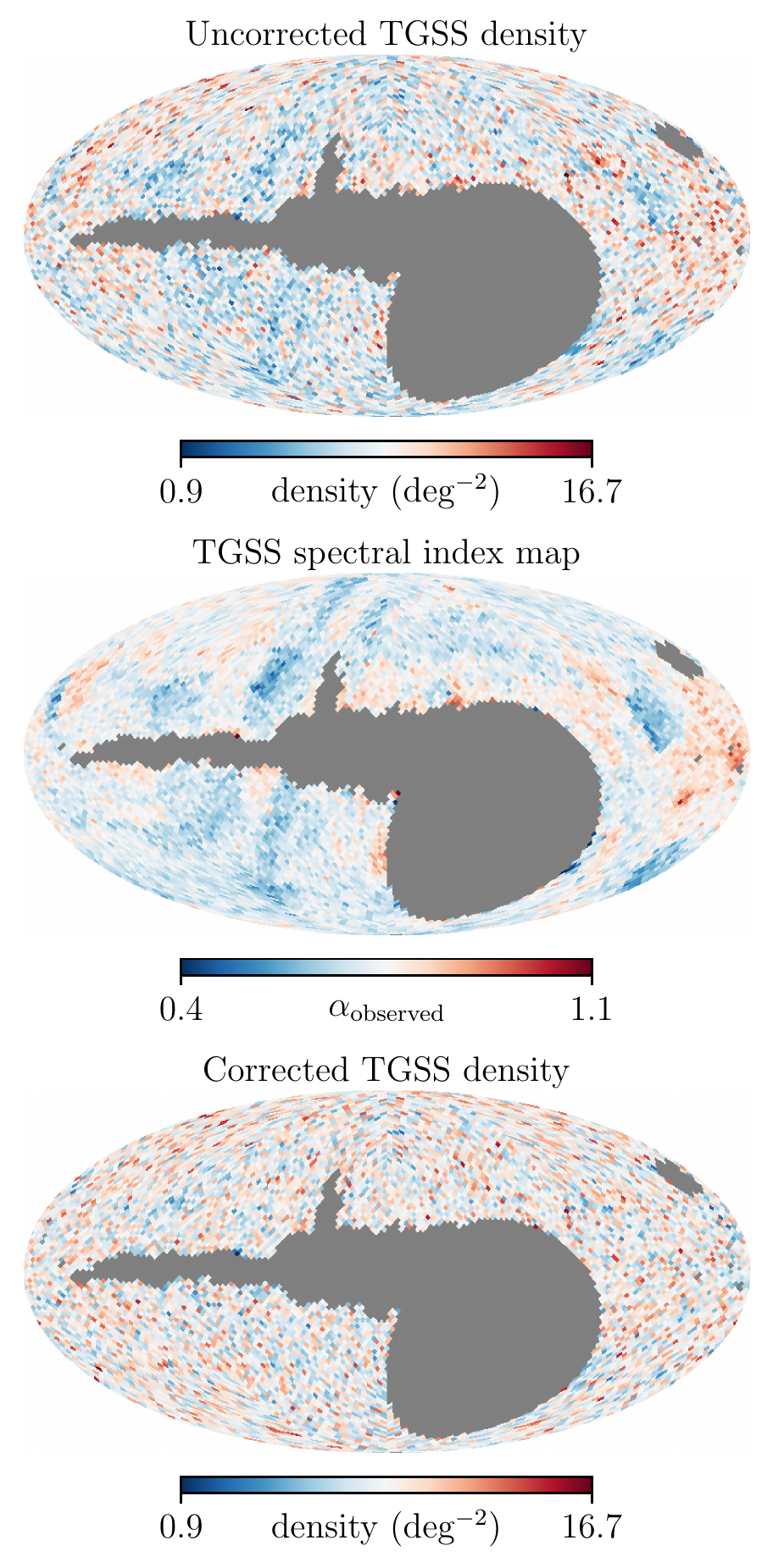}
    \caption{Top: TGSS before flux calibration correction. Middle: spectral index map. Bottom: TGSS after flux calibration correction.}
    \label{fig:tgss_demo}
\end{figure}

To estimate the effect of the flux calibration systematic, we exploit the nearly full NVSS membership of TGSS above $-40\arcdeg$ declination to determine the flux calibration correction by leveraging the large scale ($\sim10\arcdeg-30\arcdeg$) nature of the calibration issue, requiring that the mean value of the true spectral index $\alpha$ be the typical value for radio AGNs at lower frequencies: $\langle\alpha\rangle\sim0.75$. Then, the correction factor for each sky pixel is:

\begin{equation} 
f_\mathrm{corr.} = \left(\frac{150}{1400}\right)^{\langle\alpha\rangle_\mathrm{obs.} - 0.75}
\label{eq: fluxcorr}
\end{equation}

\noindent where $\langle\alpha\rangle_\mathrm{obs.}$ is the observed mean spectral index in the sky pixel, calculated using the TGSS and NVSS flux densities of each source in the pixel. We used $N_\mathrm{side}=32$ for the 0.27 million sources with corrected flux densities above 100~mJy in the masked map described below. The $\alpha$ map created using this method shows a clear correlation with the large-scale systematics visible in the TGSS sky map (Figure~\ref{fig:tgss_demo}), and correction using Equation~(\ref{eq: fluxcorr}) largely mitigates them. 

We mask the map of TGSS sources with corrected flux densities using the same procedure applied to the NVSS, cutting out sky pixels with a mean 408~MHz brightness temperature greater than 50~K. Without any flux calibration correction, we find a dipole amplitude of $6.5\times10^{-2}$, in agreement with previous estimates. After the flux calibration correction, however, the amplitude drops by a factor of $\sim 3$ to $2.2\times10^{-2}$, much closer to our NVSS/WISE result as well as the WENSS, SUMSS, and NVSS results from \citet{2021A&A...653A...9S}. This demonstrates that the apparently larger dipole in the TGSS can almost entirely be attributed to its known flux calibration issues, a result that is consistent with previous work demonstrating that position-dependent flux calibration errors in the TGSS on $\sim10\arcdeg$ scales can induce spurious power in the $l\lesssim20$ multipoles \citep{2019ApJ...887..175T}.

We quantified the effect of the RMS noise-dependent source density systematic by making sky pixel bins of source density with respect to noise, requiring 100 pixels per bin for $N_\mathrm{side}=32$. We find that source density and RMS noise show a correlation of $-0.68$ ($p\sim0$). Correcting the source density map for this correlation, the TGSS dipole becomes $\mathcal{D}=1.4\times10^{-2}$, completely consistent with the NVSS/WISE dipole. We note that dedicated studies have examined the mean spectral index of sources shared between the NVSS and TGSS, finding $\langle\alpha\rangle\sim0.78$--0.79 above the 100~mJy flux density cut we employ here \citep[respectively]{2018MNRAS.474.5008D, 2019RAA....19...96T}. Our result is not particularly sensitive to choice of fiducial mean spectral index, with these slightly steeper values yielding $\mathcal{D}=1.3\times10^{-2}$. Moreover, varying the assumed value of $\langle\alpha\rangle$ from 0.70 to 0.80 in increments of 0.01, the value of $\mathcal{D}$ stays between $1.3\times10^{-2}$ and $1.5\times10^{-2}$.


Given this result, and the consistency of the NVSS and WISE dipoles, which span nearly five orders of magnitude in frequency, we find no evidence for a frequency-dependent dipole. Indeed, the most likely scenario is that the known flux density calibration and noise systematics of the TGSS are wholly responsible for its apparently much larger dipole compared to the NVSS.

\subsection{AllWISE}
Soon after our CatWISE2020-based result was published \citep{2021ApJ...908L..51S}, \citet{2021Univ....7..107S} published an analysis of  0.28~million AGNs selected from the AllWISE-based AGN catalog of \citet{2015ApJS..221...12S}, finding a dipole amplitude of $\sim3\times10^{-2}$, i.e. a factor of $\sim2$ larger than found by \cite{2021ApJ...908L..51S} and here. The AllWISE catalog uses data exclusively from the cryogenic and post-cryogenic phases of the WISE mission prior to the hibernation period, so does not, unlike CatWISE2020, use data from the NEOWISE~Reactivation mission. Consequently, AllWISE is considerably shallower and less uniform, requiring \citet{2021Univ....7..107S} to use a much harder magnitude cut of $W1 < 15$, which greatly reduces the sample size. Nonetheless, this discrepancy warrants investigation.

We first checked that the dipole estimator used in \citet{2021Univ....7..107S} gives results consistent with ours by reproducing their result using the \citet{2015ApJS..221...12S} catalog and applying the same cuts, $12<W1<15$ and $|b|\geq 15\arcdeg$. Because of the small sample size (0.28~million), we used $N_\mathrm{side}=32$, which gives a monopole of 31 sources per sky pixel. We find $\mathcal{D}=3.1\times10^{-2}$, with a direction within $12\arcdeg$ of that found by \citet{2021Univ....7..107S}. This offset may be attributable to slight differences in the sample resulting from how the sky is masked when working with sky pixels versus source vectors. 

\begin{figure}
    \includegraphics[width=\columnwidth]{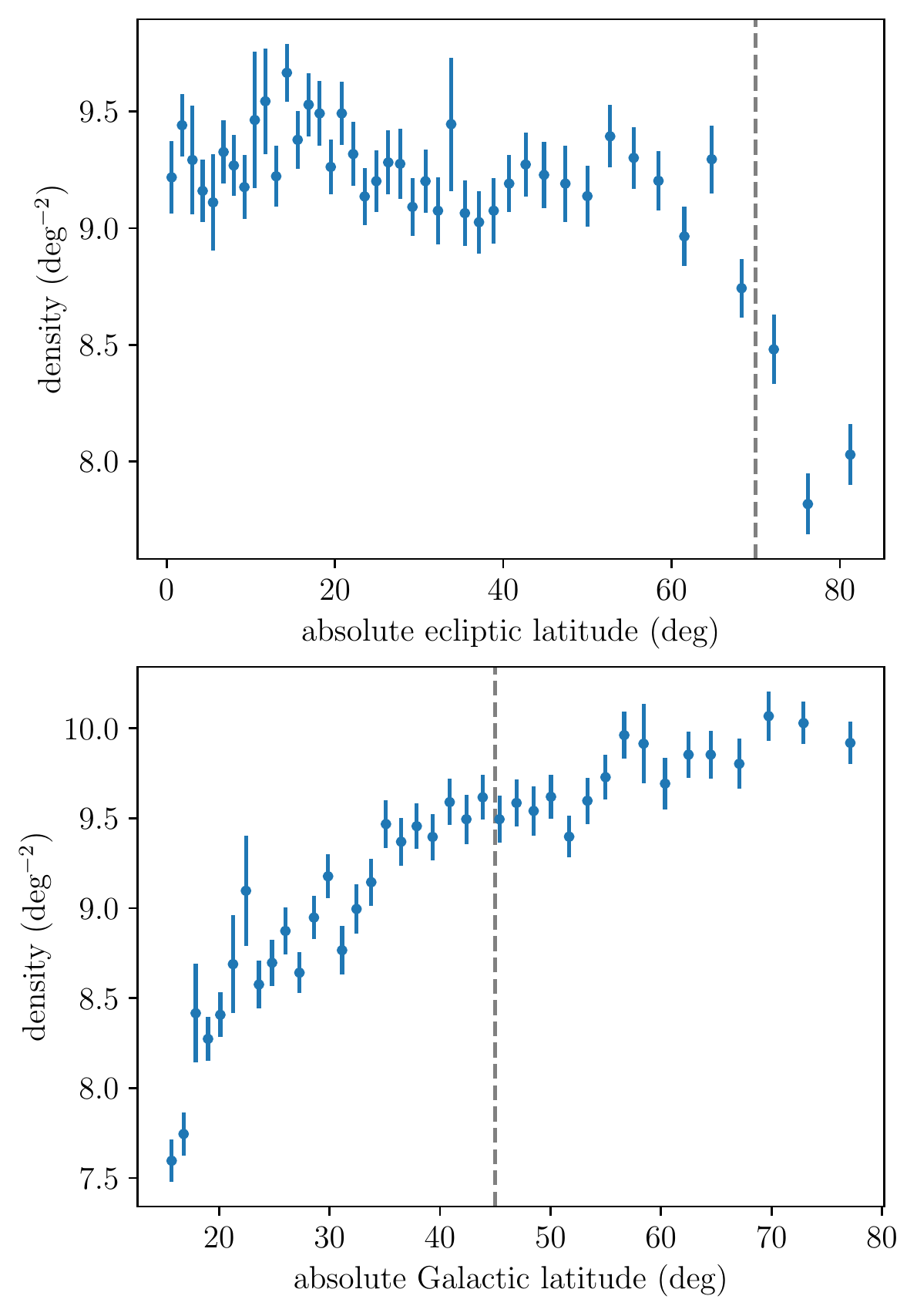}
    \caption{Top: drop in source density near ecliptic poles in the AllWISE-based AGN catalog used by \citet{2021Univ....7..107S}. Bottom: trend with Galactic latitude. The dashed lines denote the cuts we employ to account for these effects.}
    \label{fig:miragn_trends}
\end{figure}

Having obtained a consistent result, we now explore systematics in this sample. The first is the presence of stripes of reduced sensitivity along certain ecliptic lines of longitude evident in Figure~1 of \citet{2021Univ....7..107S}. We identify 4 stripes at ecliptic longitudes $10\arcdeg < \lambda < 14\arcdeg$, $238\arcdeg < \lambda < 242\arcdeg$, $313\arcdeg < \lambda < 317\arcdeg$, and $342\arcdeg < \lambda < 346\arcdeg$. Masking these, the dipole amplitude drops to $\mathcal{D}=2.6\times10^{-2}$. We also find that, although the sample does not exhibit the linear ecliptic latitude trend of the deeper CatWISE2020-based sample, it does show a steep drop off in source density at the ecliptic poles beyond $|\beta| \gtrsim 70\arcdeg$ (Figure~\ref{fig:miragn_trends}, top). Making this cut mitigates source density dependence on ecliptic latitude, although it has a minor effect on the dipole, reducing it to $\mathcal{D}=2.5\times10^{-2}$. Finally, there is a clear downward trend in source density at lower Galactic latitudes in the \citet{2015ApJS..221...12S} sample (Figure~\ref{fig:miragn_trends}, bottom), which is likely due to differences in the source detection algorithms employed for producing the AllWISE and CatWISE2020 catalogs. The latter is based on source detections from the unWISE catalog \citep{2019ApJS..240...30S} which performs better in crowded regions such as the Galactic plane and ecliptic poles. Consequently, while a cut of $|b| > 30$ was sufficient to remove dependence on Galactic latitude in the CatWISE2020 sample ($\chi^2/\mathrm{df}=1.2$), a cut of $|b| > 45$ is required for the AllWISE-based sample ($\chi^2/\mathrm{df}=1.8$). This leaves 0.11~million sources with $\mathcal{D}=1.2\times10^{-2}$, pointing $74\arcdeg$ from the CMB dipole and $85\arcdeg$ from the CatWISE2020 dipole.

We conclude that, once survey and source detection systematics have been accounted for, the AllWISE-based AGN sample employed in \citet{2021Univ....7..107S} does not exhibit a significantly larger dipole than the CatWISE2020-based quasar sample that we have used in this work, which has $\sim15$ times as many sources. As a check, we added the de-striping mask to the CatWISE2020 sample, but found it has a negligible impact, resulting in an amplitude $\mathcal{D}=1.5\times10^{-2}$ and a shift in direction of $3\arcdeg$.

\subsection{VLA Sky Survey and Rapid ASKAP Continuum Survey}
Recently, \citet{2022ApJ...931L..14D} presented an analysis of the radio galaxy dipole in the 3~GHz VLA~Sky~Survey \citep[VLASS;][]{2020PASP..132c5001L} combined with the 0.9~GHz Rapid~ASKAP~Continuum Survey \citep[RACS;][]{2020PASA...37...48M}, claiming agreement with the kinematic expectation. We examine this result below but first note two issues. \citet{2022ApJ...931L..14D} determines consistency with the kinematic expectation by fitting the dipole amplitude and direction, correcting for bias where needed, and employs bootstrap resampling to determine uncertainties. The dipole of the joint VLASS$+$RACS catalog is found to be consistent with the kinematic expectation, although it is acknowledged that their result is not inconsistent with \citet{2021ApJ...908L..51S} either. In the present paper, the effect of counting statistics and masking, along with any possible bias in the estimator, is fully accounted for in the null sky simulations, so the formal significance of our results is unaffected. This is a major motivation for our approach, as opposed to attempting to determine uncertainties and bias factors on the best-fit dipole amplitudes and directions, and working backwards to determine agreement with the kinematic expectation \citep[as was done in][and elsewhere]{2022ApJ...931L..14D}. Moreover, combining radio catalogs made at different frequencies is inherently problematic, as for a given flux limit a higher frequency catalog will preferentially select flat-spectrum sources, while a lower frequency catalog will preferentially select steep-spectrum sources, so the assumption of a characteristic, aggregate spectral index in the joint catalog may not be valid. This, and other observational systematics that may vary between the two catalogs, is the reason why we did not join the NVSS with other catalogs, such as SUMSS \citep[as was done in][]{2017MNRAS.471.1045C}.

Nonetheless, the \citet{2022ApJ...931L..14D} result deserves examination. Using the VLASS and RACS catalogs, we reproduced their joint catalog of 711,450 sources. Taking $\alpha=0.98$ and $x=1.0$ as in that work, we find a dipole amplitude of $\mathcal{D}=0.49\times10^{-2}$, in agreement with the kinematic expectation. The direction is $(l,b)=(284\arcdeg, 43\arcdeg)$, offset from the direction found by \citet{2022ApJ...931L..14D} using a similar estimator, but within $15\arcdeg$ of the CMB dipole. 

However, if the VLASS and RACS catalogs are jointly consistent with the kinematic expectation, then they should also be individually consistent, accounting for their source counts and sky masks. The advantage of our methodology is that it is straightforward to test this, by simulating skies according to the kinematic expectation, masking them identically, and then determining how consistent the found dipole directions and offsets are with expectations. For VLASS, we find $\mathcal{D}=1.0\times10^{-2}$, $80\arcdeg$ from the CMB dipole, and a $p$-value of 0.07 using $10^6$ simulations, in tension with the kinematic expectation. For RACS, which has larger overall sky coverage (63\% vs.\ 56\% for VLASS), we find $\mathcal{D}=1.5\times10^{-2}$, $42\arcdeg$ from the CMB dipole, and a $p$-value of 0.003, which is inconsistent with the kinematic expectation. 

Thus, while the joint catalog appears to be consistent with the kinematic interpretation of the CMB dipole, at least one of the individual catalogs is not, and it is possible that the ostensible overall consistency with the kinematic expectation is a coincidence of the particular distributions of sources in each catalog. Indeed, \citet{2022ApJ...931L..14D} notes that, when tested individually, the VLASS dipole points towards the south equatorial pole ($\sim75\arcdeg$ from the CMB dipole) with an amplitude corresponding to 683~km~s$^{-1}$, while the RACS dipole is closer in direction to the CMB dipole ($\sim42\arcdeg$) but with an amplitude corresponding to 644~km~s$^{-1}$.

\bibliography{manuscript}
\bibliographystyle{aasjournal}

\end{document}